\documentclass[10pt,a4paper]{iopart}
\usepackage[latin1]{inputenc}
\usepackage[T1]{fontenc}
\usepackage{iopams}
\usepackage{amsopn}
\usepackage{graphicx}
\usepackage{cite}
\usepackage{siunitx}
\usepackage{tikz}
\usepackage{array}
\usepackage{subfigure}
\usepackage{longtable}
\usepackage{listings}

\newcommand{\LTspice}{LTspice IV}
\DeclareMathOperator{\varRe}{Re}
\DeclareMathOperator{\varIm}{Im}
\newcommand{\ju}{\ensuremath{\mathrm{j}}}

\begin{document}
\sisetup{
	retain-explicit-plus = true
}
\title{Circuit models and SPICE macro-models for quantum Hall effect devices}
\author{Massimo Ortolano\textsuperscript{1,2} and Luca Callegaro\textsuperscript{2},
		\\[\medskipamount]  
		     {\small \textsuperscript{1} Dipartimento di Elettronica e Telecomunicazioni, Politecnico di Torino } \\
        {\small Corso Duca degli Abruzzi 24, 10129 Torino, Italy} \\
        {\small \textsuperscript{2}INRIM - Istituto Nazionale di Ricerca Metrologica} \\
        {\small Strada delle Cacce, 91 - 10135 Torino, Italy} \\
}

\begin{abstract}
Quantum Hall effect (QHE) devices are a pillar of modern quantum electrical metrology. Electrical networks including one or more QHE elements can be used as quantum resistance and impedance standards. The analysis of these networks allows metrologists to evaluate the effect of the inevitable parasitic parameters on their performance as standards. This paper presents a systematic analysis of the various circuit models for QHE elements proposed in the literature, and the development of a new model. This last model is particularly suited to be employed with the analogue electronic circuit simulator SPICE. The SPICE macro-model and examples of SPICE simulations, validated by comparison with the corresponding analytical solution and/or experimental data, are provided. 
\end{abstract}

\maketitle
\section{Introduction}
The quantum Hall effect (QHE) has been the basis of resistance metrology for several years in DC~\cite{Delahaye:2003,Jeckelmann:2003,Poirier:2011} and more recently in the AC regime~\cite{Overney:2006,Schurr:2007,Ahlers:2009,Hernandez:2014}. In DC, networks of several QHE elements have been developed to realize quantum Hall array resistance standards (QHARS)~\cite{Piquemal:1999,Poirier:2002,Bounouh:2003,Poirier:2004,Hein:2004,Oe:2008,Oe:2011,Woszczyna:2012,Oe:2013}, intrinsically-referenced voltage dividers \cite{Domae:2012} and Wheatstone bridges \cite{Schopfer:2007}. In AC, Schurr \emph{et al.}~\cite{Schurr:2009} included two QHE elements in a quadrature bridge to realize the unit of capacitance.

In 1988, Ricketts and Kemeny (RK)~\cite{Ricketts:1988} proposed a circuit model suitable for the symbolic analysis of arbitrary electrical networks containing QHE elements, in DC or AC. The predictions of the RK model were also verified experimentally~\cite{Ricketts:1988,Jeffery:1995}. Other circuit models were developed in later years~\cite{Hartland:1995,Chua:1999b,Jeffery:1995,Cage:1998a,Sosso:1999,Sosso:2001,Schurr:2006,Schurr:2014} and a general method of analysis based on the indefinite admittance matrix has been recently proposed~\cite{Ortolano:2012}. 

The circuit models proposed in the literature are better suited for the symbolic analysis of networks of QHE elements; the matrix method can be employed both for symbolic and numerical analyses, typically with the aid of a computer algebra system or a numerical computing environment. Nowadays, however, the numerical analysis of electrical circuits is commonly carried out by employing analogue electronic circuit simulation tools, like SPICE~\cite{SpiceHome} and its derivatives. SPICE can perform simulations of linear and non-linear networks in time and frequency domains, and noise analysis. Hence, it could be a very useful tool for the simulation of electrical networks containing QHE elements. Unfortunately, even though the above cited circuit models can be directly coded in SPICE, several issues arise when trying to run a simulation. The aim of this work is to address these issues and to present a working SPICE macro-model for the simulation of networks of QHE elements.

In section~\ref{sec:review}, we review the existing circuit models for QHE elements and present a new one. In section~\ref{sec:spice_modelling}, we discuss the issues of SPICE modelling and provide a SPICE macro-model for an 8-terminal QHE element, which is based on the new circuit model. In section~\ref{sec:examples}, three simulation examples are worked out in detail: i) a DC analysis taking into account parasitic resistances; ii) an AC analysis of a network containing a QHE element and a capacitor; iii) a noise analysis. The simulation of example i) is compared with an analytical result; those of ii) and iii) are compared with analytical and experimental results. \ref{sec:yam} reports the full derivation of the new circuit model.

\section{Brief review of QHE circuit models}
\label{sec:review}
The development of circuit models for the Hall effect is not a recent topic. For models related to the classical Hall effect see, e.g.,\cite{Garg:1965,Arnold:1982,Popovic:1985,Salim:1992,Salim:1995}; in this section, instead, we shall briefly review the existing circuit models for \emph{ideal} QHE elements. 

\begin{figure}
  \centering
  \includegraphics[clip=]{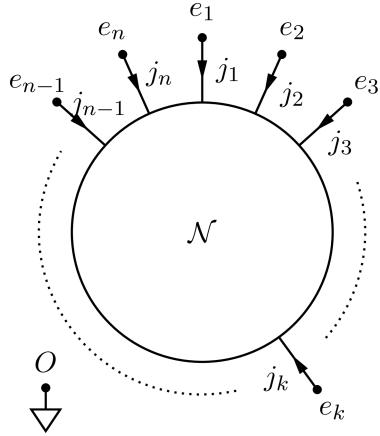}
  \caption{An $n$-terminal element $\mathcal{N}$ represented with reference polarities and directions: terminals are numbered in a clockwise direction; terminal voltages are measured with respect to an arbitrary reference point $O$ (\emph{datum node}); terminal currents flowing into $\mathcal{N}$ are considered positive. Voltages and currents can be functions of time.}\label{fig:nterminal_element}
\end{figure}

In circuit theory, an $n$-\emph{terminal element} is a black-box which can interact with its surroundings solely through $n$ conductors called \emph{terminals} (figure~\ref{fig:nterminal_element}). A basic assumption of circuit theory is that an \mbox{$n$-terminal} element is completely characterized by its external behaviour, that is, by the set of admissible terminal voltage-current pairs (action at distance is excluded)~\cite{Chua:1980,Willems:2010}. Two elements are thus \emph{equivalent} from the point of view of circuit theory if they have the same external behaviour.

An electric circuit composed of several interconnected elements can be itself turned into an $n$-terminal element by connecting terminals to $n$ selected nodes of the circuit. If a circuit, viewed as an $n$-terminal element, is equivalent to a certain $n$-terminal element $\mathcal{N}$, it is called a \emph{circuit model} (or \emph{equivalent circuit}) for \textbf{$\mathcal{N}$}.

According to their external behaviour, ideal QHE elements can be classified in two categories. In the following, we shall call\footnote{The terminology has been inspired by the two circuit models shown in the first row of table~\ref{tab:circuit_models}.} an \emph{ideal clockwise (cw) \mbox{$n$-terminal} QHE element} with Hall resistance $R_\textup{H}>0$, an element whose terminal voltages and currents are related by the equations
\begin{equation}\label{eq:cw_element}
R_\textup{H}\,j_m = e_m-e_{m-1}\,,\qquad m=1,\ldots, n, 
\end{equation}
with the convention $e_0\equiv e_n$, and we shall call an \emph{ideal counterclockwise (ccw) $n$-terminal QHE element}, an element whose terminal voltages and currents are related by the equations
\begin{equation}\label{eq:ccw_element}
R_\textup{H}\,j_m = e_m-e_{m+1}\,,\qquad m=1,\ldots, n,
\end{equation}
with the convention $e_{n+1}\equiv e_1$. Ideal QHE elements are \emph{memoryless}, \emph{passive} (actually, they dissipate power) and \emph{nonreciprocal}\footnote{They are memoryless because the sets of equations~\eref{eq:cw_element} and~\eref{eq:ccw_element} relate terminal voltages and currents at the same time instant. For definitions of passivity and reciprocity see e.g.~\cite[Ch.~2]{Belevitch:1968}.}.

The sets of equations~\eref{eq:cw_element} and~\eref{eq:ccw_element} are, indeed, a consequence of the QHE phenomenology, but can also be derived theoretically on the basis of the Landauer-B\"uttiker formalism (see, e.g., \cite[Sec.~16.3]{Ihn:2010} and references therein).

A real QHE device operating at appropriate values of temperature and magnetic flux density can be modelled, with a certain approximation, either as an ideal cw or an ideal ccw element depending on the orientation of the magnetic field applied to the device and on the type of the majority charge carriers. The Hall resistance is quantized, $R_\textup{H} = R_\textup{K}/i$, where $R_\textup{K} = h/e^2$ is the von Klitzing constant and $i$ is a positive integer called \emph{plateau index}. In QHE devices realized with GaAs technology, the plateau index of interest for resistance metrology is $i=2$, so $R_\textup{H} = R_\textup{K}/2$. The recommended value of the von Klitzing constant is $R_\textup{K} = \SI{25812.8074434(84)}{\ohm}$, which yields $R_\textup{H} = R_\textup{K}/2 = \SI{12906.4037217(42)}{\ohm}$~\cite{CODATA:2010}. The conventional value adopted internationally for resistance metrology is $R_{\textup{K-90}} = \SI{25812.807}{\ohm\of{90}}$, which yields $R_{\textup{H-90}} = R_{\textup{K-90}}/2 = \SI{12906.4035}{\ohm\of{90}}$. In section~\ref{sec:examples}, we shall use this last value for the Hall resistance of the QHE elements in the example circuits.

\begin{table*}
\centering
\caption{List of known circuit models for QHE elements. The first column reports references to the original works. The second and third columns report, with uniform symbology, the corresponding circuit models for ideal cw and ccw 4-terminal QHE elements: terminal voltages and currents are chosen according to figure~\ref{fig:nterminal_element}; $e_{i,j} = e_i-e_j$; $r = R_\textup{H}/2$; diamond symbols represent controlled sources. In the last column, outlines of the circuit models for a cw 6-terminal element are sketched (see the given references for details).}
\label{tab:circuit_models}
\begin{tabular}{m{32mm}>{\centering\arraybackslash}m{41mm}>{\centering\arraybackslash}m{41mm}>{\centering\arraybackslash}m{40mm}}
\br
\multicolumn{1}{c}{Reference} & 4-terminal cw model & 4-terminal ccw model & Outline of the 6-terminal model (cw) \\ \mr
\footnotesize 1.\rule{1ex}{0ex}Ricketts-Kemeny~\cite{Ricketts:1988} & \includegraphics{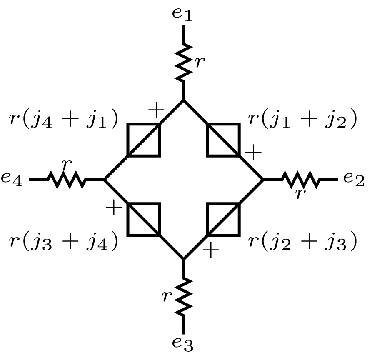} & \includegraphics{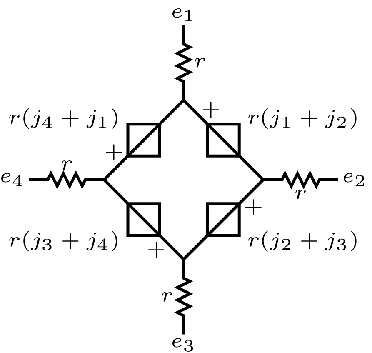} & \includegraphics{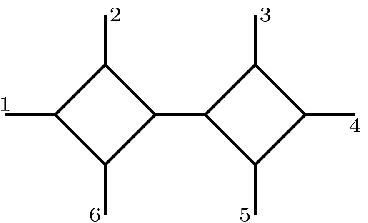} \\
\footnotesize 2.\rule{1ex}{0ex}Hartland \emph{et al.}~\cite{Hartland:1995,Chua:1999b} & \includegraphics{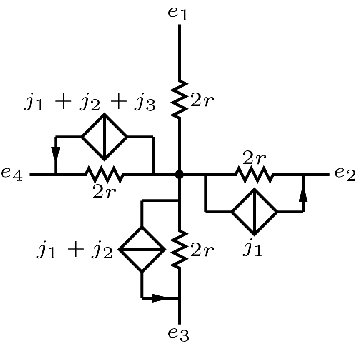} & \includegraphics{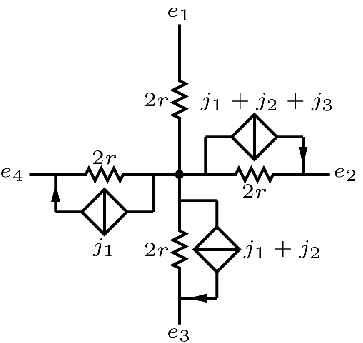} & No generalization has been directly proposed by the authors \\
\footnotesize 3.\rule{1ex}{0ex}Jeffery \emph{et al.}~\cite{Jeffery:1995,Cage:1998a} & \includegraphics{fig_rk_cw} & \includegraphics{fig_rk_ccw} & \includegraphics{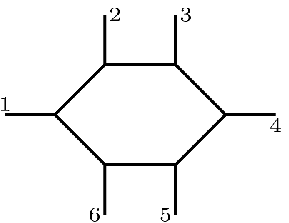} \\
\footnotesize 4.\rule{1ex}{0ex}Sosso-Capra~\cite{Sosso:1999,Sosso:2001} & \includegraphics{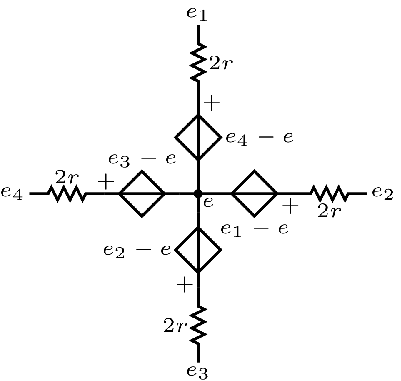} & \includegraphics{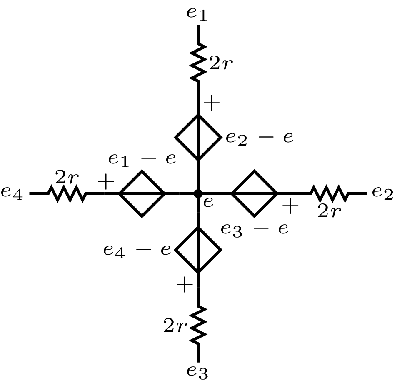} & \includegraphics{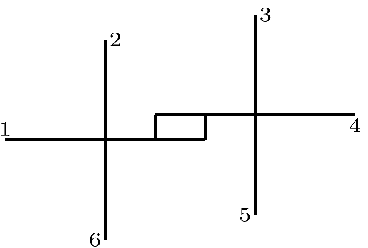} \\
\footnotesize 5.\rule{1ex}{0ex}Schurr \emph{et al.}~\cite{Schurr:2006,Schurr:2014} & \includegraphics{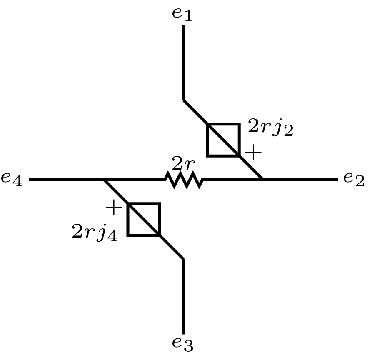} & \includegraphics{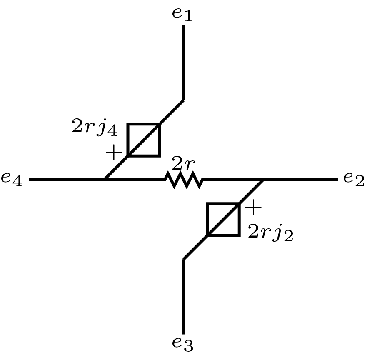} & \includegraphics{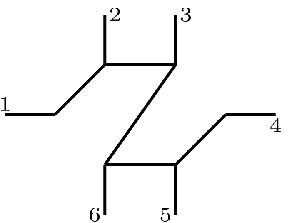} \\
\footnotesize 6.\rule{1ex}{0ex}This work\par (\ref{sec:yam}) & \includegraphics{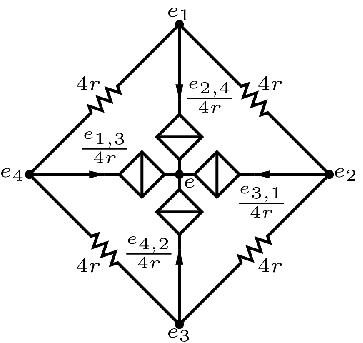} & \includegraphics{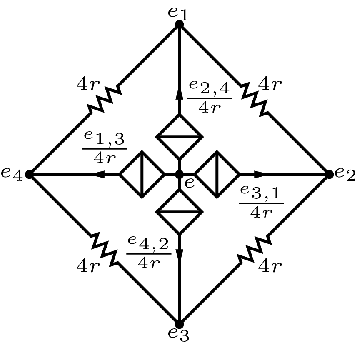} & \includegraphics{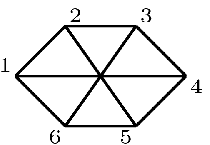} \\
\br
\end{tabular}
\end{table*}

Table~\ref{tab:circuit_models} reports known circuit models listed in order of appearance in the literature: the first row reports the well-known circuit of Ricketts and Kemeny~\cite{Ricketts:1988}; the last one, a new circuit model whose derivation is given in~\ref{sec:yam}. For each referenced work, the second and third columns show the circuit models corresponding to the ideal cw and ccw \mbox{4-terminal} QHE elements (the cited works might not provide both); in order to see how these models can be extended to elements with more than 4 terminals, the fourth column shows the outlines of the circuit models corresponding to a cw 6-terminal element. All circuit models contain only resistors (because of memorylessness and passivity) and controlled sources (because of nonreciprocity); none of them are related to the actual device physics. 

We remark that the sets of equations~\eref{eq:cw_element} and~\eref{eq:ccw_element} are all that is needed to analyse circuits containing ideal QHE elements~\cite{Ortolano:2012}, and that the external behaviour of all the models listed in table~\ref{tab:circuit_models} is exactly described by those equations. Nonetheless, a circuit model can be a useful basis for taking into account nonidealities (e.g.\ nonzero longitudinal resistances~\cite{Cage:1998a,Schurr:2006}, parasitic elements in AC regime~\cite{Cage:1999a}) or to develop a SPICE macro-model, as we shall see in the next section. 

\section{SPICE modelling}
\label{sec:spice_modelling}
In SPICE a circuit is represented by a list of statements (\emph{netlist}), written according to a certain specific syntax~\cite{SpiceHome,Steer:2007}, which describes how the elements that compose the circuit are interconnected. A SPICE macro-model (or \emph{subcircuit netlist}) is a netlist with a designated name that can be treated as any other SPICE element to compose larger circuits in a hierarchical way.   
Circuits composed exclusively of resistors and controlled sources can be directly translated into SPICE macro-models. Given that all the models reported in table~\ref{tab:circuit_models} are equivalent from the point of view of circuit theory, and that many more possessing this property can be conceived, how can we select one with the purpose of writing a SPICE macro-model? 

In reality, SPICE modelling requires at least two additional properties, (\ref{property:1}) and (\ref{property:2}) below, but one more, (\ref{property:3}), would be useful: 
\begin{enumerate}
\item\label{property:1} \emph{No loops of voltage sources}. SPICE requires a circuit that does not contain loops composed only of ideal voltage sources (independent or controlled). In fact, violating this requirement would lead to two scenarios: either Kirchhoff's voltage law would be violated or the current crossing the loop would be indeterminate and the circuit would not have a unique solution. In both cases, SPICE would not be able to solve the circuit.
\item\label{property:2} \emph{No cut sets of current sources}. This property is the dual of~(\ref{property:1}). If, in a circuit, there is a node where only current sources join, either Kirchhoff's current law would be violated or the node potential would be indeterminate (floating node).
\item\label{property:3} \emph{Non-dissipative/generative sources}. As stated in section~\ref{sec:review}, QHE elements are passive elements that dissipate power. One property that could then be required by the circuit model is that all the power is dissipated in the resistors and that the overall power absorbed or delivered by the controlled sources is zero. This property would also allow SPICE to predict the thermal noise generated by the QHE elements correctly (see~\ref{sec:yam} for more details on this). Even though this property is not strictly required, and even though SPICE has a few limitations in noise analysis (e.g.\ it is not able to analyse correlations directly and a workaround is needed~\cite{McAndrew:2005}), it might be useful to have a circuit model which is as complete as possible.
\end{enumerate}
By inspection, it is easy to verify that models 1 and 3 of table~\ref{tab:circuit_models} violate property~(\ref{property:1}) and that model 6 violates property~(\ref{property:2}); property~(\ref{property:1}) can also be violated by model 5 if a voltage source is connected between terminals 1 and 2, or 3 and 4. In addition, it is straightforward to verify that models 2,4 and 5 violate property~(\ref{property:3})\footnote{For instance, for model 5 (cw), the power delivered by the controlled sources is $p=2r(j_1 j_2+j_3j_4)$, which is generally not zero.}. 

Models 2,4 and 5 cannot be easily modified to satisfy property~(\ref{property:3}) and will not be considered further. Models 1 and 3 can be modified to satisfy property~(\ref{property:1}) either by adding a small series resistance to the loop of voltage sources~\cite{Rashid:2006} or by removing one of the controlled sources in the loop (this does not alter the model's behaviour). However, none of the above two remedies is completely satisfactory: the former because the effect of the additional resistance on the simulation accuracy cannot be easily predicted in the case of many interconnected QHE elements; the latter because it destroys the symmetry of the model, making it more difficult to develop possible refinements to take account of nonidealities. For model 6, it is worth noting that the potential $e$ of the floating node can be fixed arbitrarily without altering the model's behaviour; this allows the satisfaction of property~(\ref{property:2}) without the need for additional elements and without destroying the circuit symmetry. Therefore, model 6 was selected to implement a SPICE macro-model.

Listings~\ref{lst:cw} and~\ref{lst:ccw} report, respectively, the SPICE macro-models for ideal cw and ccw \mbox{8-terminal} QHE elements\footnote{These macro-models should work out of the box, or with just slight modifications, with any modern SPICE simulator that accepts subcircuits with parameters. In this work, we have chiefly used \LTspice\ from Linear Technology Corporation~\cite{LTspice}, but tests have been carried out also with the open source simulator Ngspice~\cite{Ngspice}, with PSpice A/D from Cadence Design Systems and with TINA-TI from Texas Instruments. The authors can provide advice on adapting the circuit models and the simulations here described to other SPICE-based analogue circuit simulators.}. These macro-models are a direct translation of the general circuit models derived in~\ref{sec:yam}. The actual number of terminals is nine: the ninth terminal, labelled C, corresponds to the node joining the controlled current sources and has to be connected to an arbitrary potential  (e.g.\ ground). The default value of $R_\textup{H}$ is \SI{1}{\ohm}, but other values can be assigned when calling the macro-model (see next section for an example).

\lstset{basicstyle=\footnotesize\ttfamily}
\begin{lstlisting}[caption={SPICE macro-model for an ideal cw 8-terminal element.},label=lst:cw]
.subckt qhe8cw 1 2 3 4 5 6 7 8 C params: RH=1
R1 1 2 {2*RH}
R2 2 3 {2*RH}
R3 3 4 {2*RH}
R4 4 5 {2*RH}
R5 5 6 {2*RH}
R6 6 7 {2*RH}
R7 7 8 {2*RH}
R8 8 1 {2*RH}
G1 1 C 2 8 {1/(2*RH)}
G2 2 C 3 1 {1/(2*RH)}
G3 3 C 4 2 {1/(2*RH)}
G4 4 C 5 3 {1/(2*RH)}
G5 5 C 6 4 {1/(2*RH)}
G6 6 C 7 5 {1/(2*RH)}
G7 7 C 8 6 {1/(2*RH)}
G8 8 C 1 7 {1/(2*RH)}
.ends
\end{lstlisting}

\begin{lstlisting}[caption={SPICE macro-model for an ideal ccw 8-terminal element.},label=lst:ccw]
.subckt qhe8ccw 1 2 3 4 5 6 7 8 C params: RH=1
R1 1 2 {2*RH}
R2 2 3 {2*RH}
R3 3 4 {2*RH}
R4 4 5 {2*RH}
R5 5 6 {2*RH}
R6 6 7 {2*RH}
R7 7 8 {2*RH}
R8 8 1 {2*RH}
G1 C 1 2 8 {1/(2*RH)}
G2 C 2 3 1 {1/(2*RH)}
G3 C 3 4 2 {1/(2*RH)}
G4 C 4 5 3 {1/(2*RH)}
G5 C 5 6 4 {1/(2*RH)}
G6 C 6 7 5 {1/(2*RH)}
G7 C 7 8 6 {1/(2*RH)}
G8 C 8 1 7 {1/(2*RH)}
.ends
\end{lstlisting}

\section{Examples}
\label{sec:examples}

\subsection{Double-series interconnection of two QHE elements}
Several QHE elements can be interconnected to obtain multiples, submultiples or fractions of the Hall resistance. Multiple-series, parallel\cite{Delahaye:1993} and bridge connections~\cite{Ortolano:2015} can be employed to reject the effect of the inevitable contact and wiring resistances. In this section, SPICE is used to analyse, in the DC regime, the effect of the parasitic resistances in a double-series connection; the result of the SPICE analysis is then compared with an analytical solution obtained with the technique described in~\cite{Ortolano:2012}. 

\begin{figure}
  \centering
  \includegraphics[clip=]{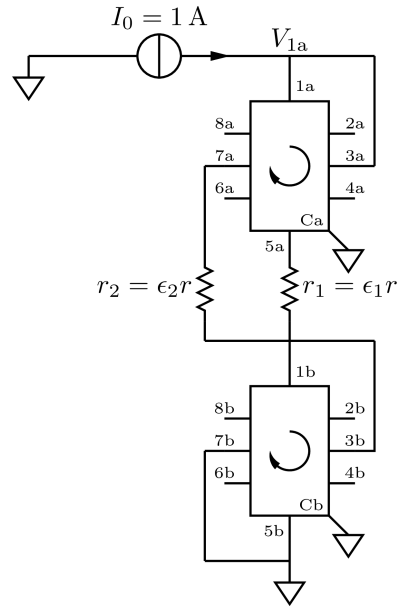}
  \caption{Circuit diagram for the DC analysis of a double-series connection. $r_1$ and $r_2$ represent the wiring and contact resistances between the two devices; other contact resistances are neglected for simplicity.}\label{fig:ds}
\end{figure}

Figure~\ref{fig:ds} shows the complete circuit diagram for the analysis of the double-series connection. For ease of analysis, only two parasitic resistances, $r_1 = \epsilon_1 r$ and $r_2 = \epsilon_2 r$ ($r = R_\textup{H}/2$), are taken into account. The resistance $R_\textup{S}^{(2)}$ of the double-series is calculated between the terminal 1a and ground, $R_\textup{S}^{(2)} = V_\textup{1a}/I_0$, where $I_0$ can be chosen of \SI{1}{\ampere} to simplify the calculation. The effect of the parasitic resistances can be quantified by the relative discrepancy
\begin{equation}\label{eq:ds_delta}
	\delta = \frac{R_\textup{S}^{(2)}-2R_\textup{H}}{2R_\textup{H}},
\end{equation}
where $2R_\textup{H}$ is the resistance of the double-series when $r_1 = r_2 = \SI{0}{\ohm}$. At the third order in $\epsilon_1$ and $\epsilon_2$, the analytical solution yields\footnote{The authors can provide the Mathematica\textregistered\ notebook used to obtain the analytical solution.}
\begin{equation}
	\delta^\textup{theo} = \frac{\epsilon_1 \epsilon_2}{16}-\frac{\epsilon_1^2 \epsilon_2+\epsilon_1 \epsilon_2^2}{64}\,.
\end{equation}

\begin{lstlisting}[caption={SPICE netlist corresponding to the circuit of figure~\ref{fig:ds}. The included file \texttt{qhe8cw.sub} should contain the SPICE macro-model of listing~\ref{lst:cw}. In this example we have set $\epsilon_1 = \num{0.15}t$ and $\epsilon_2 = \num{0.35}t$, where $t$ is a dimensionless scaling parameter. The dot command \texttt{.op} declares that SPICE should perform a DC operating point analysis. The dot command \texttt{.meas} (\LTspice\ specific) performs the calculation of the parameter $\delta$ as defined by~\eref{eq:ds_delta}.},label=lst:ds]
QHE double-series circuit simulation
* Includes the macro-model
.inc qhe8cw.sub
* Definition of circuit parameters
.param RH=12906.4035 r={RH/2}
.param t=0.01
.param r1={0.15*t*r} r2={0.35*t*r}
* Circuit netlist
XU1 1A 2A 1A 4A 5A 6A 7A 8A 0 qhe8cw 
+params: RH={RH}
XU2 1B 2B 3B 4B 0 6B 0 8B 0 qhe8cw 
+params: RH={RH}
I0 0 1A 1
r1 5A 1B {r1}
r2 7A 3B {r2}
* Analysis directives
.op
.meas op delta param (V(1A)-2*RH)/(2*RH)
.end
\end{lstlisting}

The SPICE netlist\footnote{Modern SPICE simulators come with handy graphical schematic editors which alleviate the chore of writing netlists. Here, however, we have preferred to provide netlists, using basic elements and directives, which can be easily converted from one SPICE dialect to another (the dialect employed here is mainly that of \LTspice).} which corresponds to the circuit of figure~\ref{fig:ds} is reported in listing~\ref{lst:ds}, with example parasitic resistances $r_1 = \num{0.15}t\,r$ and $r_2 = \num{0.35}t\,r$, where we have introduced a scaling parameter $t$ to check the round-off errors due to floating point arithmetic and numerical algorithms. From~\eref{eq:ds_delta}, $\delta$ is expected to scale as $t^2$. 

Table~\ref{tab:ds_results} reports a comparison, for different values of $t$, between the values of $\delta^\textup{theo}$ obtained from~\eref{eq:ds_delta} and those of $\delta^\textup{sim}$ obtained from the SPICE simulation of listing~\ref{lst:ds}. The third column $\delta^\textup{sim}$ of table~\ref{tab:ds_results} can be considered representative of the results obtainable with common SPICE simulators: as can be seen from the reported results, when $\delta$ falls below $\num{e-10}$, $\delta^\textup{sim}$ diverges from $\delta^\textup{theo}$. The last column $\delta^\textup{sim}$ (alt) reports, instead, the results obtained with the \LTspice's alternate solver, which, according to the \LTspice\ manual, has reduced round-off errors. In any case, whatever the simulator employed in this kind of analysis, we suggest to write all parasitic resistances as functions of a scaling parameter to keep under control round-off errors.

\begin{table}
\centering
\caption{Comparison, for the circuit of figure~\ref{fig:ds} with $r_1 = \num{0.15}t\,r$ and $r_2 = \num{0.35}t\,r$, between the relative discrepancy $\delta^\textup{theo}$, obtained from~\eref{eq:ds_delta}, and $\delta^\textup{sim}$, obtained by simulating with SPICE the corresponding netlist of listing~\ref{lst:ds}. The third column $\delta^\textup{sim}$ reports the results obtained with the \LTspice's default solver; similar results are obtained with other SPICE-based simulators. The column $\delta^\textup{sim}$ (alt) reports the results obtained with the \LTspice's alternate solver with reduced round-off errors.}
\label{tab:ds_results}
\begin{tabular}{@{}S[table-format = e+2]S[table-format = +1.2e+2]@{\hspace{1ex}}S[table-format = +1.2e+2]@{\hspace{1ex}}S[table-format = +1.2e+2]@{}}
\br
{$t$} & {$\delta^\textup{theo}$} & {$\delta^\textup{sim}$} & {$\delta^\textup{sim}$ (alt)} \\ 
\mr
e-1 & 3.24e-5 & 3.24e-5 & 3.24e-5 \\
e-2 & 3.28e-7 & 3.28e-7 & 3.28e-7 \\
e-3 & 3.28e-9 & 3.28e-9 & 3.28e-9 \\
e-4 & 3.28e-11 & 5.62e-11 & 3.28e-11 \\
e-5 & 3.28e-13 & -1.70e-10 & 3.22e-13 \\
\br
\end{tabular}
\end{table}

\subsection{QHE gyrator}
\label{sec:qhe_gyrator}
\begin{figure}
  \centering
  \includegraphics[clip=]{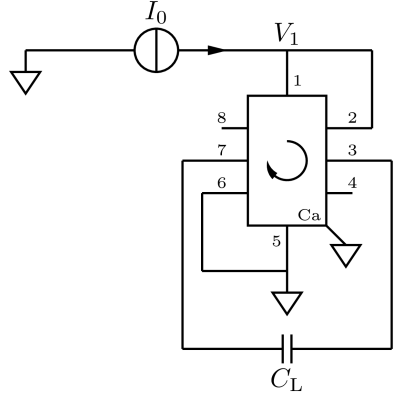}
  \caption{Circuit diagram for the AC analysis of a QHE element used as gyrator. The impedance of interest is $Z(\omega) = V_1/I_0$.}\label{fig:gyrator}
\end{figure}

QHE elements, like their classical counterparts, can be employed as \emph{gyrators}~\cite{Tellegen:1948,Viola:2014}. At the angular frequency $\omega$, the circuit of figure~\ref{fig:gyrator}, which includes the load capacitor $C_\textup{L}$ with negative reactance $X_\textup{L} = -(\omega C_\textup{L})^{-1}$, realizes between terminal 1 and ground an impedance $Z(\omega) = V_1/I_0 = R(\omega)+\ju X(\omega)$ having \emph{positive} reactance, that is, $X(\omega) > 0$ for any $\omega$. Therefore, at any given frequency, the circuit of figure~\ref{fig:gyrator} behaves like an $RL$ two-terminal element (the values of $R$ and $L$ are frequency-dependent).

Network analysis yields
\begin{equation}\label{eq:gyrator}
 Z(\omega) = \frac{R_\textup{H}}{1+\omega^2 C_\textup{L}^2 R_\textup{H}^2} \left[(1 + 2 \omega^2 C_\textup{L}^2 R_\textup{H}^2) + \ju\omega C_\textup{L} R_\textup{H} \right],
\end{equation}
for which $X(\omega) = \varIm Z(\omega)>0$.

\begin{figure}
  \centering
  \includegraphics[clip=]{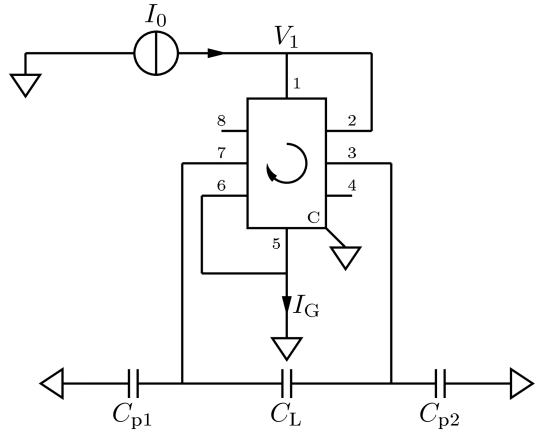}
  \caption{Circuit diagram for the AC analysis of a QHE element used as gyrator in the experimental set-up described in the text. $C_\textup{p1}$ and $C_\textup{p2}$ represent the cable capacitances. The LCR bridge measures the impedance $Z^\textup{meas} = V_1/I_\textup{G}$, which is different from the impedance $Z(\omega)$ defined in figure~\ref{fig:gyrator}.}\label{fig:gyrator_lcr}
\end{figure}
 
The behaviour of such an unconventional QHE circuit was checked experimentally with an 8-terminal GaAs-AlGaAs Hall bar working at the temperature $T=\SI{1.6}{\kelvin}$ and plateau index $i=2$. Terminals 3 and 7 were connected to a variable capacitance box; terminals pairs $(1,2)$ and $(5,6)$ were connected in double-series configuration to an LCR meter (Agilent mod.~4284A) which performed the impedance measurements. The measurement frequency $f = \omega/(2\pi)$ was chosen at $\SI{1233}{\hertz}$, so that the reactance of a \SI{10}{\nano\farad} capacitor is $X_\textup{L} \approx -R_\textup{H}$. 

However, the above described experimental set-up is not accurately modelled by the circuit of figure~\ref{fig:gyrator} and by the corresponding equation~\eref{eq:gyrator}, because in figure~\ref{fig:gyrator} the cable capacitances and the real operation of the LCR meter are not taken into account (see~\cite{Hernandez:2014} for an example about the effect of the cable capacitances in AC QHE measurements). The experimental set-up is instead better modelled by the circuit of figure~\ref{fig:gyrator_lcr}, where $C_\textup{p1}$ and $C_\textup{p2}$ represent the cable capacitances, and where the transimpedance $Z^\textup{meas} = V_1/I_\textup{G}$ represents the impedance actually measured by the LCR meter, that is, the ratio of the meter's high-side voltage to the meter's low-side current.

Even though the circuit of figure~\ref{fig:gyrator_lcr} can be solved analytically, this kind of analysis is not straightforward, and SPICE simulation can provide a quicker response. Listing~\ref{lst:gyrator_lcr} reports the SPICE netlist for the AC analysis of the circuit of figure~\ref{fig:gyrator_lcr}, for different values of $C_\textup{L}$ at the fixed frequency of $\SI{1233}{\hertz}$. The given values for $C_\textup{L}$ are those used in the experiment: \SIlist{0; 1; 3; 7; 10; 15; 20; 30}{\nano\farad}. The values of the parasitic capacitances reported in the listing, $C_\textup{p1} = C_\textup{p2} = \SI{300}{\pico\farad}$, are just rough estimates and not the result of a measurement.   

\begin{lstlisting}[caption={SPICE netlist corresponding to the circuit of figure~\ref{fig:gyrator_lcr}.  The \texttt{.step} directive is used to automatically run simulations with different values of $C_\textup{L}$; the \texttt{.ac} directive declares that SPICE should perform an AC sweep analysis, here of just one frequency point; the \texttt{.probe} directive is used to save node voltages and branch currents for further analysis, plots etc.},label=lst:gyrator_lcr]
QHE gyrator simulation
* Includes the macro-model
.inc qhe8cw.sub
* Definition of circuit parameters
.param RH=12906.4035
.step param CL list 0 1n 3n 7n 10n 15n 20n 30n
* Circuit netlist
XU1 1 1 3 4 5 5 7 8 0 qhe8cw 
+params: RH={RH}
I0 0 1 AC 1
CL 3 7 {CL}
Cp1 7 0 300p
Cp2 3 0 300p
VG 5 0 0
* Analysis directives
.ac lin 1 1233 1233
.probe V(1) I(VG)
.end
\end{lstlisting}

\begin{figure}
\centering
\includegraphics[width=\columnwidth,clip=]{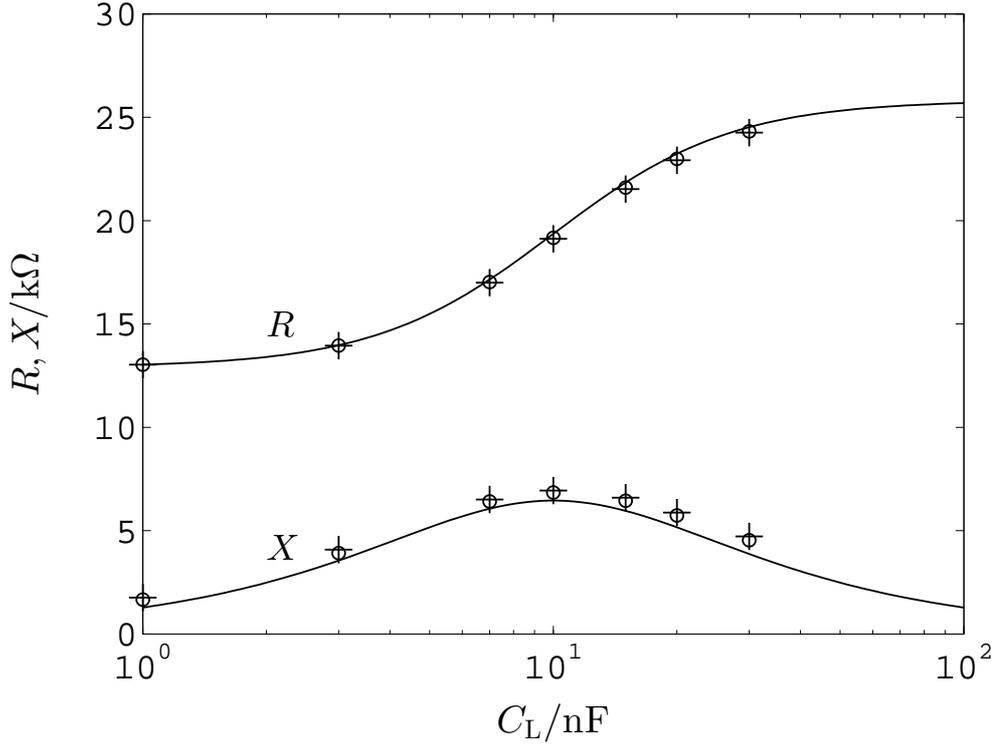}
\caption{QHE gyrator: comparison of results from experiment ($+$), simulation ($\circ$) and equation~\eref{eq:gyrator} (solid line), for different values of the load capacitance $C_\textup{L}$ at the fixed frequency of \SI{1233}{\hertz}.}
\label{fig:gyrator_comparison}
\end{figure}

Results are reported in figure~\ref{fig:gyrator_comparison}, where the measured values of $R(C_\textup{L})$ and $X(C_\textup{L})$ at the frequency of \SI{1233}{\hertz} are compared with the values obtained from the SPICE simulation of listing~\ref{lst:gyrator_lcr} and with those obtained from~\eref{eq:gyrator}. The agreement between the measured values and those obtained from the SPICE simulation is within \SI{1}{\percent}. Further model refinements can be easily implemented in the SPICE simulation.

\subsection{Noise analysis}
Noise is a fundamental limit to the resolution of a measuring system. Noise analysis is therefore a useful tool to predict the resolution achievable in a measurement. SPICE has built-in noise models for resistors and semiconductor devices which allow a complete noise analysis of circuits, taking into account all the main noise sources. In the case of circuits containing QHE elements, the noise analysis can include the thermal noise generated by these elements and the noise generated by possible auxiliary elements, such as voltage or current sources, detectors etc.

\begin{figure}
\centering
\includegraphics[clip=]{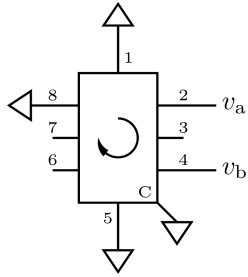}
\caption{Example circuit for noise analysis. The objective is to evaluate the cross-spectral density function $S_\textup{ab}(f)$ between the voltages $v_\textup{a}$ and $v_\textup{b}$.}
\label{fig:noise_a}
\end{figure}

In this section we present the noise analysis of the circuit of figure~\ref{fig:noise_a}, for which the thermal noise properties have been investigated both theoretically and experimentally in~\cite{Callegaro:2013}. The one-sided cross-spectral density function $S_\textup{ab}(f)$ between the voltages $v_\textup{a}$ and $v_\textup{b}$ due to thermal noise is $S_\textup{ab}(f) = 2k_\textup{B}T R_\textup{H}$~\cite{Callegaro:2013}, where $k_\textup{B}$ is the Boltzmann's constant and $T$ is the thermodynamic temperature of the QHE element.

\begin{figure}
\centering
\includegraphics[clip=]{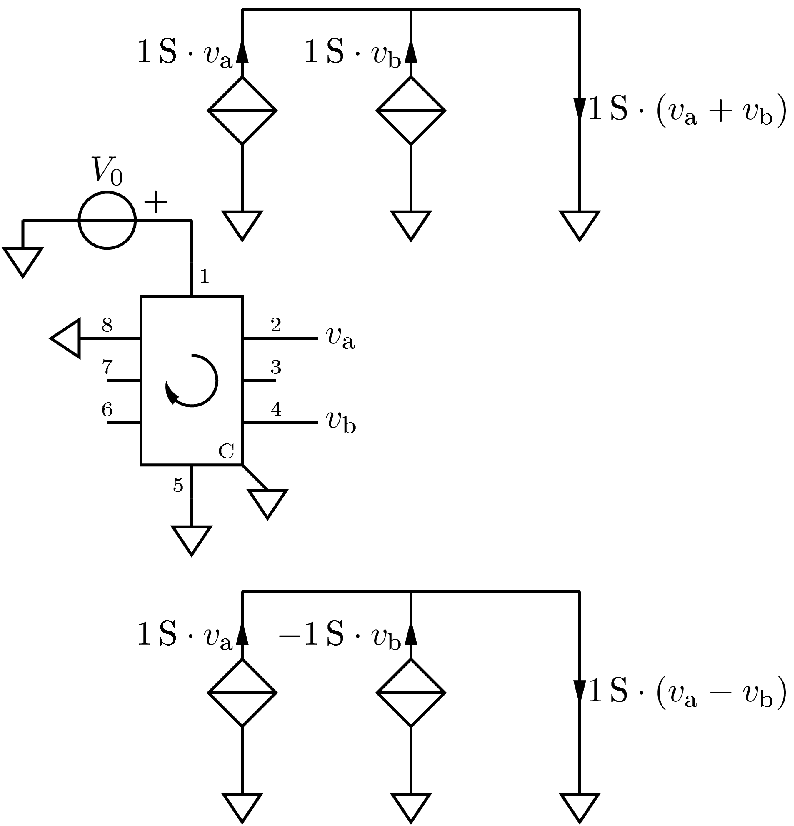}
\caption{The circuit of figure~\ref{fig:noise_a} with the additional elements needed for the SPICE evaluation of $S_\textup{ab}(f)$.}
\label{fig:noise_b}
\end{figure}

Figure~\ref{fig:noise_b} shows the circuit needed to evaluate $S_\textup{ab}(f)$ with SPICE. Since SPICE cannot evaluate directly cross-spectral density functions, but only (auto-) spectral density functions, a trick should be employed~\cite{McAndrew:2005}. Four voltage controlled current sources, with transconductances of \SI{1}{\siemens}, generate two currents, one proportional to the sum of $v_\textup{a}$ and $v_\textup{b}$, the other proportional to their difference; it can be shown that
\begin{equation}
 S_\textup{ab}(f) = \frac{1}{4}[S_{+}(f)-S_{-}(f)],
\end{equation}
where $S_{+}(f)$ is the spectral density function of $v_\textup{a}+v_\textup{b}$ and $S_{-}(f)$ is the spectral density function of $v_\textup{a}-v_\textup{b}$. Both $S_{+}(f)$ and $S_{-}(f)$ can be directly evaluated by SPICE. The additional, noiseless, voltage source $V_0$ is actually ineffective in this analysis, but it is required by SPICE to have a fictitious input. 

Listing~\ref{lst:noise} reports the SPICE netlist for the noise analysis described above.  The subcircuit \verb|probe_correlation| is used to calculate the sum and the difference of $v_\textup{a}$ and $v_\textup{b}$ in two successive steps: when the parameter \texttt{gb} is set to 1 the output node (\texttt{OUT}) of \verb|probe_correlation| yields the sum of the two voltages; when \texttt{gb} is set to -1, the difference is instead obtained. The simulation temperature is set to \SI{2}{\kelvin}. The calculation of $S_\textup{ab}(f)$ or of the ratio $S_\textup{ab}(f)/(4k_\textup{B}R_\textup{H}T)$ can be done automatically by the SPICE graphical post-processor, but we shall not dwell on the details here. The result obtained from the simulation coincides with the theoretical one.

\begin{lstlisting}[caption={SPICE netlist corresponding to the circuit of figure~\ref{fig:noise_b}.},label=lst:noise]
QHE noise simulation
* Includes the macro-model
.inc qhe8cw.sub
* Defines a subcircuit for calculating
* the sum and the difference of two
* voltages
.subckt probe_correlation A B OUT
G3 0 J A 0 1
G4 0 J B 0 {gb}
V2 0 J 0
H2 OUT 0 V2 1
.ends probe_correlation
* Definition of circuit parameters
.param RH=12906.4035
.temp -271.15
.step param gb list -1 1
* Circuit netlist
XU1 1 2 3 4 0 6 7 0 0 qhe8cw 
+params: RH={RH}
V0 1 0 0
XX1 2 4 OUT probe_correlation
* Analysis directives
.noise V(OUT) V0 dec 10 100 10k
.end
\end{lstlisting} 

\section{Conclusions}
In this paper we reviewed the modelling of the ideal quantum Hall effect element as a $n$-terminal electrical network. The unique set of network equations can be represented with different equivalent circuit models published in literature. An analysis of these models shows that they are unsuitable to be directly coded in circuit simulation software such as SPICE, because they generate singular representations or because of errors in the noise simulation. A new QHE circuit model, specifically designed to be implemented in SPICE, is proposed. This paper provides several examples of electrical circuits including QHE elements, of their coding in SPICE and hints for proper simulation runs. For each case, the simulation outcome is compared with the results from analytical modelling and also, in two cases, with experimental data.

\section*{Acknowledgments}
The authors are grateful to Franz Josef Ahlers and J\"urgen Schurr from the Physikalisch-Technische Bundesanstalt (PTB), Braunschweig (Germany), for their collaboration in the experiment described in section~\ref{sec:qhe_gyrator}, which was carried out at the Electrical Quantum Metrology Department of the PTB. 

The authors would also like to thank Helmut Sennewald and the LTspice IV users' group~\cite{LTspiceUG} for useful suggestions on the usage of LTspice IV.

\appendix
\section{Yet another model}
\label{sec:yam}

\subsection{Model derivation}
In this section we derive model 6 of table~\ref{tab:circuit_models} for an ideal $n$-terminal QHE element on the basis of the analysis method described in~\cite{Ortolano:2012} and briefly reviewed below. In the derivation, sinusoidal regime is assumed: voltages and currents are to be understood as voltage and current phasors and are denoted with capital letters. Labelling of terminals and reference directions are again that of figure~\ref{fig:nterminal_element}.

The sets of equations~\eref{eq:cw_element} and~\eref{eq:ccw_element} can be rewritten in matrix form as~\cite{Ortolano:2012}
\begin{equation}\label{eq:iam_general}
\bi{J} = \bar{\bi{Y}}_\textup{i}\bi{E}\,,
\end{equation}
where $\bi{J} = (J_1,\ldots,J_n)^\textup{T}$ (the superscript T denotes transposition) and $\bi{E} = (E_1,\ldots,E_n)^\textup{T}$ are column vectors, and where the $n\times n$ \emph{indefinite admittance matrix} (see~\cite{Shekel:1954} for a definition) $\bar{\bi{Y}}_\textup{i}$ is either
\begin{equation}\label{eq:iam_cw}
\bar{\bi{Y}}_\textup{i,cw} = \frac{1}{R_\textup{H}}\left(\begin{array}{rrrrrr}
1 &0 &\cdots  &0 &-1 \\
-1 &1 &0 &\cdots &0 \\
 &\ddots &\ddots &\ddots &\vdots \\
0 & \cdots & -1 & 1 &0   \\
0 & 0 & \cdots &-1 &1
\end{array}\right)\,,
\end{equation}
in the case of an ideal cw $n$-terminal element, or
\begin{equation}\label{eq:iam_ccw}
\bar{\bi{Y}}_\textup{i,ccw} =  \frac{1}{R_\textup{H}}\left(\begin{array}{rrrrrr}
1 &-1 &0 &\cdots &0 \\
0 &1 &-1 &\ddots &0 \\
\vdots &\ddots &\ddots &\ddots &\vdots \\
0 &\cdots &0 &1 &-1   \\
-1 &0 &\cdots &0 &1
\end{array}\right)\,,
\end{equation}
in the case of an ideal ccw $n$-terminal element. 

The two above matrices are real matrices. A basic theorem of algebra~\cite[Ch.~1]{Roman:2010} states that every matrix can be decomposed into the sum of a symmetric part and an antisymmetric one:
\begin{equation}\label{eq:matrix_decomposition}
\bar{\bi{Y}}_\textup{i} = \bar{\bi{Y}}_\textup{i}^\textup{S}+\bar{\bi{Y}}_\textup{i}^\textup{A},
\end{equation}
with
\begin{equation}
\bar{\bi{Y}}_\textup{i}^\textup{S} = \frac{1}{2}(\bar{\bi{Y}}_\textup{i}+\bar{\bi{Y}}_\textup{i}^\textup{T})\quad\text{and}\quad\bar{\bi{Y}}_\textup{i}^\textup{A} = \frac{1}{2}(\bar{\bi{Y}}_\textup{i}-\bar{\bi{Y}}_\textup{i}^\textup{T})\,.
\end{equation}
Since, by definition, $(\bar{\bi{Y}}_\textup{i}^\textup{S})_{kl} = (\bar{\bi{Y}}_\textup{i}^\textup{S})_{lk}$, the matrix $\bar{\bi{Y}}_\textup{i}^\textup{S}$ is associated to a reciprocal network~\cite[Ch.~16]{Desoer:1969}; in addition, this associated network can be realized with just resistors because $\bar{\bi{Y}}_\textup{i}^\textup{S}$ is also real. The antisymmetric part $\bar{\bi{Y}}_\textup{i}^\textup{A}$ is instead associated to a nonreciprocal network that can be realized with voltage controlled current sources. For the matrices in~\eref{eq:iam_cw} and~\eref{eq:iam_ccw} the symmetric and antisymmetric parts are
\begin{eqnarray}\label{eq:iam_sym}
\bar{\bi{Y}}_\textup{i,cw}^\textup{S} &= \bar{\bi{Y}}_\textup{i,ccw}^\textup{S} \nonumber \\
&= \frac{1}{R_\textup{H}}\left(\begin{array}{rrrrrr}
1 &-\frac{1}{2} &0 &\cdots &-\frac{1}{2} \\
-\frac{1}{2} &1 &-\frac{1}{2}&0 &\cdots \\
 & \ddots &\ddots &\ddots &\vdots \\
0 &\cdots &-\frac{1}{2} &1 &-\frac{1}{2}\\
-\frac{1}{2} &0 &\cdots &-\frac{1}{2} &1
\end{array}\right)\,,
\end{eqnarray}
\begin{equation}\label{eq:iam_cw_anti}
\bar{\bi{Y}}_\textup{i,cw}^\textup{A} = \frac{1}{R_\textup{H}}\left(\begin{array}{rrrrrr}
0 &\frac{1}{2} &0  &\cdots &-\frac{1}{2} \\
-\frac{1}{2} &0 &\frac{1}{2} &\ddots &0 \\
 &\ddots &\ddots &\ddots &\vdots \\
0 &\ldots &-\frac{1}{2} &0 &\frac{1}{2}  \\
\frac{1}{2} &0 &\cdots &-\frac{1}{2} &0
\end{array}\right)
\end{equation}
and
\begin{eqnarray}\label{eq:iam_ccw_anti}
\bar{\bi{Y}}_\textup{i,ccw}^\textup{A} &= -\bar{\bi{Y}}_\textup{i,cw}^\textup{A} \nonumber \\
&= \frac{1}{R_\textup{H}}\left(\begin{array}{rrrrrr}
0 &-\frac{1}{2} &0  &\cdots &\frac{1}{2} \\
\frac{1}{2} &0 &-\frac{1}{2} &\ddots &0 \\
 &\ddots &\ddots &\ddots &\vdots \\
0 &\ldots &\frac{1}{2} &0 &-\frac{1}{2}  \\
-\frac{1}{2} &0 &\cdots &\frac{1}{2} &0
\end{array}\right)\,.
\end{eqnarray}

From~\eref{eq:iam_general}, taking into account~\eref{eq:matrix_decomposition}, we obtain that the terminal currents can be decomposed into the sum of two contributions, one associated to $\bar{\bi{Y}}_\textup{i}^\textup{S}$ and one to $\bar{\bi{Y}}_\textup{i}^\textup{A}$:
\begin{equation}
\bi{J} = \bi{J}^\textup{S}+\bi{J}^\textup{A}\,,
\end{equation}
with
\begin{equation}\label{eq:current}
\bi{J}^\textup{S} = \bar{\bi{Y}}_\textup{i}^\textup{S}\bi{E}\quad\text{and}\quad\bi{J}^\textup{A} = \bar{\bi{Y}}_\textup{i}^\textup{A}\bi{E}\,.
\end{equation}
This means that $\bar{\bi{Y}}_\textup{i}^\textup{S}$ and $\bar{\bi{Y}}_\textup{i}^\textup{A}$ can be considered associated to two electrical networks connected in parallel. 

\begin{figure}
\centering
\includegraphics[scale=1]{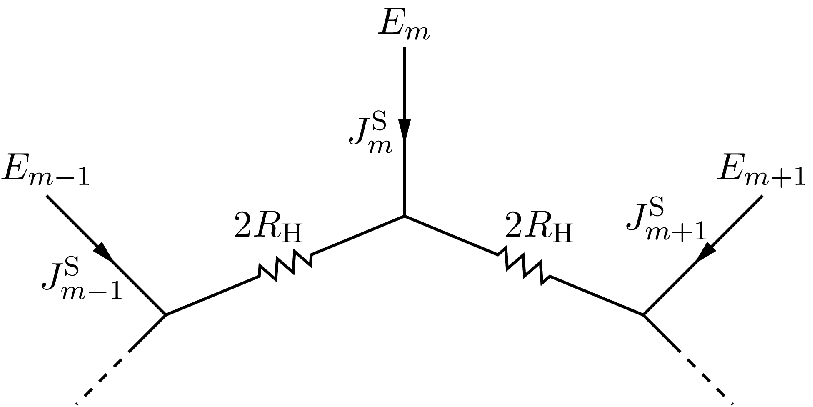}
\caption{Portion of the circuit model associated with the symmetric part $\bar{\bi{Y}}_\textup{i}^\textup{S}$ of $\bar{\bi{Y}}_\textup{i}$.}
\label{fig:model_sym}
\end{figure}

Let $\bi{J}^\textup{S}=(J_1^\textup{S},\ldots,J_n^\textup{S})^\textup{T}$: from~\eref{eq:current} and~\eref{eq:iam_sym} the current at the $m$th terminal is ($m=1,\ldots,n$; $E_0\equiv E_n$ and $E_{n+1}\equiv E_1$)
\begin{eqnarray}
J_m^\textup{S} &= \frac{E_m-E_{m-1}/2+E_{m+1}/2}{R_\textup{H}}\,, \\
&= \frac{E_m-E_{m-1}}{2R_\textup{H}}+\frac{E_m-E_{m+1}}{2R_\textup{H}}\,.
\end{eqnarray}
The last equation above corresponds to a network where each terminal is connected to its adjacent terminals through a resistor with resistance $2R_\textup{H}$, as shown in figure~\ref{fig:model_sym}. This can also be obtained directly from the properties of the indefinite admittance matrices~\cite[Ch.~2, Sec.~2.2]{Chen:1991}.

\begin{figure}
\centering
\includegraphics[scale=1]{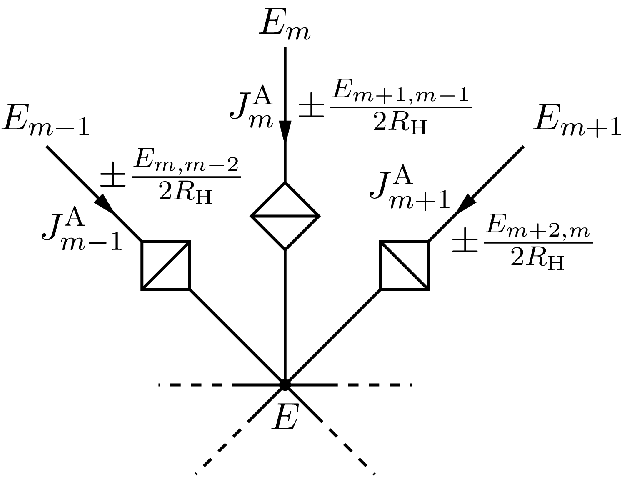}
\caption{Portion of the circuit model associated with the antisymmetric part $\bar{\bi{Y}}_\textup{i}^\textup{A}$ of $\bar{\bi{Y}}_\textup{i}$; $E_{i,j} = E_i-E_j$; for the voltage controlled current sources, the plus sign holds for a cw element and the minus for a ccw one; $E$ is the potential of an additional internal node.}
\label{fig:model_anti}
\end{figure}

Now, let $\bi{J}^\textup{A}=(J_1^\textup{A},\ldots,J_n^\textup{A})^\textup{T}$: from~\eref{eq:current}, \eref{eq:iam_cw_anti} and~\eref{eq:iam_ccw_anti}, the current at the $m$th terminal is
\begin{equation}
J_m^\textup{A} = \pm\frac{E_{m+1}-E_{m-1}}{2R_\textup{H}}\,,
\end{equation}
where the plus sign holds for a cw element and the minus for a ccw one. This equation can be obtained with a voltage controlled current source, driven by the voltage difference $E_{m+1}-E_{m-1}$, which draws from the $m$th terminal the current $\pm(E_{m+1}-E_{m-1})/(2R_\textup{H})$ injecting it into an additional internal node $E$ (figure~\ref{fig:model_anti}). 

\begin{figure}
\centering
\includegraphics[scale=1]{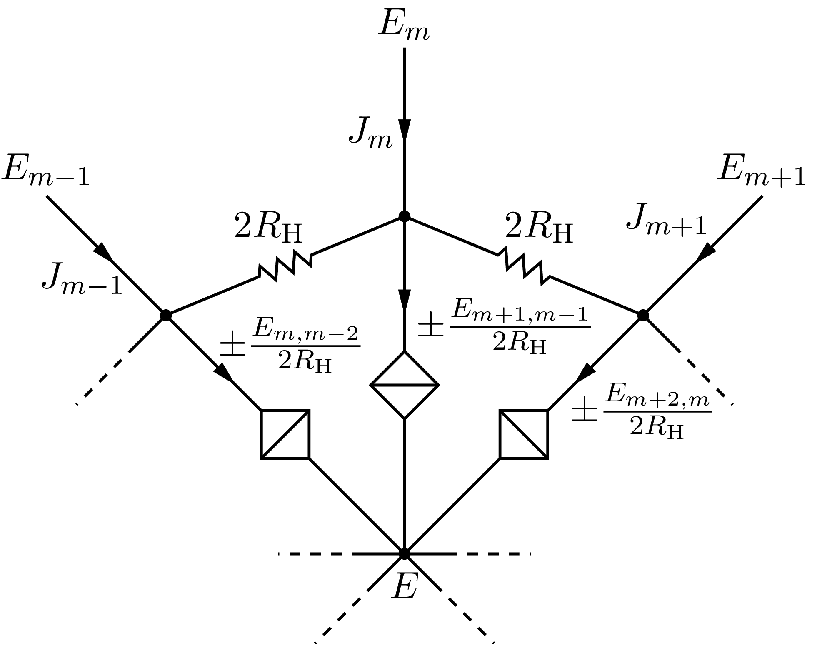}
\caption{Portion of the final circuit model obtained by combining in parallel the two circuits of figures~\ref{fig:model_sym} and~\ref{fig:model_anti}.}
\label{fig:model}
\end{figure}

Finally, combining in parallel the two networks, one obtains the circuit model of figure~\ref{fig:model}. This circuit model obviously satisfies property~(\ref{property:1}); property~(\ref{property:2}) can be satisfied by fixing the potential $E$ of the central node of figure~\ref{fig:model} to an arbitrary value.

\subsection{Power}
To prove that the circuit model of figure~\ref{fig:model} satisfies property~(\ref{property:3}), let us recall that, in sinusoidal regime, the average power entering an $n$-terminal element is given by~\cite{Shekel:1954,Desoer:1969}
\begin{eqnarray}
P_\textup{av} &= \frac{1}{2}\varRe\left(\sum_{m=1}^n J_m^* E_m\right) = \frac{1}{2}\varRe(\bi{J}^*\bi{E})\,, \\
&= \frac{1}{4}(\bi{J}^*\bi{E}+\bi{E}^*\bi{J})\,,\label{eq:power_def}
\end{eqnarray}
where the operator $\varRe$ denotes the real part and the asterisk $*$ denotes the conjugate transpose. Substituting~\eref{eq:iam_general} in~\eref{eq:power_def}, and taking into account that $\bar{\bi{Y}}_\textup{i}$ is real, yields 
\begin{eqnarray}
P_\textup{av} &= \frac{1}{4}\bi{E}^*(\bar{\bi{Y}}_\textup{i}^*+\bar{\bi{Y}}_\textup{i})\bi{E}\,, \\
&= \frac{1}{4}\bi{E}^*(\bar{\bi{Y}}_\textup{i}^\textup{T}+\bar{\bi{Y}}_\textup{i})\bi{E}\,, \\
&= \frac{1}{2}\bi{E}^*\bar{\bi{Y}}_\textup{i}^\textup{S}\bi{E}\,. \label{eq:power}
\end{eqnarray}
Equation~\eref{eq:power} says that the power dissipated in the ideal QHE element is related only to the symmetric part of the corresponding indefinite admittance matrix. Thus, no power is delivered or absorbed by the circuit associated with the antisymmetric part, and property~(\ref{property:3}) is satisfied.

\subsection{Noise modelling}
When SPICE performs the noise analysis of a circuit, it automatically assigns noise sources to certain circuit elements, in accordance with known noise models. The insertion of arbitrary noise sources is not permitted. In the case of the macro-models of listings~\ref{lst:cw} and~\ref{lst:ccw}, SPICE will assign noise sources to the resistors only: the controlled sources are considered noiseless. To each resistor, SPICE assign a noise source which generates the corresponding thermal (equilibrium) noise at the simulation temperature (specified by the SPICE parameter \texttt{TNOM}, \SI{27}{\celsius} by default). Therefore, to prove that the satisfaction of property~(\ref{property:3}) allows SPICE to predict the equilibrium noise generated by the QHE elements correctly, we need only to prove that the equilibrium noise of a QHE element coincides with that of the resistive network associated with $\bar{\bi{Y}}_\textup{i}^\textup{S}$.

The equilibrium noise properties of a general passive multiterminal network were derived by Twiss~\cite{Twiss:1955} from thermodynamic considerations. Later, B\"{u}ttiker derived an equivalent result for multiterminal conductors from a quantum scattering theory~\cite{Buttiker:1990,Buttiker:1992}\footnote{The two approaches are complementary: the derivation of Twiss can be considered analogous to that of Nyquist~\cite{Nyquist:1928} for the Johnson-Nyquist noise generated by a resistor; the derivation of B\"{u}ttiker can be considered analogous to that of Landauer~\cite{Landauer:1989}.}. These results were verified experimentally by the authors in~\cite{Callegaro:2013}.

For the present purposes, the result given by Twiss is in a more convenient form. This result can be stated as follows~\cite{Twiss:1955}: consider a passive $n$-terminal element and assume that all the terminals are grounded and one, say terminal $n$ without loss of generality, is considered as reference; then, the one-sided cross-spectral density function~\cite{Priestley:1981} $S_{j_p,j_q}(f)$ between the noise currents $j_p$ and $j_q$ at the terminals $p=1,\ldots,n-1$ and $q=1,\ldots,n-1$ is given by
\begin{equation}
S_{j_p,j_q}(f) = 2k_\textup{B} T(Y_{pq}+Y_{qp}^*)\,,
\end{equation}
where $k_\textup{B}$ is the Boltzmann's constant, $T$ is the thermodynamic temperature and $\bi{Y} = (Y_{pq})$ is the short-circuit $(n-1)\times(n-1)$ admittance matrix obtained by considering the $n$-terminal element as an $(n-1)$-port with ports defined between the terminals $1,\ldots,n-1$ and the reference node $n$. 

The short-circuit admittance matrix can be obtained by the indefinite admittance matrix associated to the $n$-terminal element by deleting the $n$th row and column~\cite{Shekel:1954}; then, for and ideal QHE element, $Y_{pq} = (\bar{\bi{Y}}_\textup{i})_{pq}$ and
\begin{eqnarray}
S_{j_p,j_q}(f) &= 2k_\textup{B} T(Y_{pq}+Y_{qp}^*), \\
&= 2k_\textup{B} T\left[\bar{(\bi{Y}}_\textup{i})_{pq}+(\bar{\bi{Y}}_\textup{i})_{qp}^*\right], \\
&= 2k_\textup{B} T\left[\bar{(\bi{Y}}_\textup{i})_{pq}+(\bar{\bi{Y}}_\textup{i})_{qp}\right], \\
&= 4k_\textup{B} T(\bar{\bi{Y}}_\textup{i}^\textup{S})_{pq}, \\
&= 2k_\textup{B} T\left[\bar{(\bi{Y}}_\textup{i}^\textup{S})_{pq}+(\bar{\bi{Y}}_\textup{i}^\textup{S})_{qp}\right].
\end{eqnarray}
Hence, and taking into account that the reference node is arbitrary, the equilibrium noise of a QHE element coincides with that of the resistive network associated with $\bar{\bi{Y}}_\textup{i}^\textup{S}$.

\section*{References}

\begin{thebibliography}{10}
\expandafter\ifx\csname url\endcsname\relax
  \def\url#1{{\tt #1}}\fi
\expandafter\ifx\csname urlprefix\endcsname\relax\def\urlprefix{URL }\fi
\providecommand{\eprint}[2][]{\url{#2}}

\bibitem{Delahaye:2003}
Delahaye F and Jeckelmann B 2003 {\em Metrologia\/} {\bf 40} 217

\bibitem{Jeckelmann:2003}
Jeckelmann B and Jeanneret B 2003 {\em Meas. Sci. Technol.\/} {\bf 14} 1229

\bibitem{Poirier:2011}
Poirier W, Schopfer F, Guignard J, Th\'{e}venot O and Gournay P 2011 {\em C. R.
  Phys.\/} {\bf 12} 347--368

\bibitem{Overney:2006}
Overney F, Jeanneret B, Jeckelmann B, Wood B~M and Schurr J 2006 {\em
  Metrologia\/} {\bf 43} 409

\bibitem{Schurr:2007}
Schurr J, Ahlers F~J, Hein G and Pierz K 2007 {\em Metrologia\/} {\bf 44}
  15--23

\bibitem{Ahlers:2009}
Ahlers F~J, Jeanneret B, Overney F, Schurr J and Wood B~M 2009 {\em
  Metrologia\/} {\bf 46} R1--R12

\bibitem{Hernandez:2014}
Hernandez C, Consejo C, Degiovanni P and Chaubet C 2014 {\em J. Appl. Phys.\/}
  {\bf 115} 123710

\bibitem{Piquemal:1999}
Piquemal F~P~M, Blanchet J, G\`{e}neves G and Andr\'{e} J~P 1999 {\em IEEE
  Trans. Instr. Meas.\/} {\bf 48} 296--300

\bibitem{Poirier:2002}
Poirier W, Bounouh A, Hayashi H, Fhima H, Piquemal F, G\`{e}neves G and
  Andr\'{e} J~P 2002 {\em J. Appl. Phys.\/} {\bf 92} 2844--2854

\bibitem{Bounouh:2003}
Bounouh A, Poirier W, Piquemal F, G\`{e}neves G and Andr\'{e} J~P 2003 {\em
  {IEEE} Trans. Instr. Meas.\/} {\bf 52} 555--558

\bibitem{Poirier:2004}
Poirier W, Bounouh A, Piquemal F and Andr\'{e} J~P 2004 {\em Metrologia\/} {\bf
  41} 285

\bibitem{Hein:2004}
Hein G, Schumacher B and Ahlers F~J 2004 Preparation of quantum {H}all effect
  device arrays {\em 2004 Conference on Precision Electromagnetic Measurements
  Digest\/} (London, UK) pp 273--274

\bibitem{Oe:2008}
Oe T, Kaneko N, Urano C, Itatani T, Ishii H and Kiryu S 2008 Development of
  quantum {H}all array resistance standards at {NMIJ} {\em Precision
  Electromagnetic Measurements Digest, 2008. CPEM 2008. Conf. on\/}
  (Broomfield, CO, USA) pp 20--21

\bibitem{Oe:2011}
Oe T, Matsuhiro K, Itatani T, Gorwadkar S, Kiryu S and Kaneko N 2011 {\em IEEE
  Trans. Instr. Meas.\/} {\bf 60} 2590--2595

\bibitem{Woszczyna:2012}
Woszczyna M, Friedemann M, Dziomba T, Weimann T and Ahlers F~J 2011 {\em Appl.
  Phys. Lett\/} {\bf 99} 022112 (pages~3)

\bibitem{Oe:2013}
Oe T, Matsuhiro K, Itatani T, Gorwadkar S, Kiryu S and Kaneko N 2013 {\em IEEE
  Trans. Instr. Meas.\/} {\bf 62} 1755--1759

\bibitem{Domae:2012}
Domae A, Oe T, Matsuhiro K, Kiryu S and Kaneko N 2012 {\em Meas. Sci.
  Technol.\/} {\bf 23} 124008

\bibitem{Schopfer:2007}
Schopfer F and Poirier W 2007 {\em J. Appl. Phys.\/} {\bf 102} 054903

\bibitem{Schurr:2009}
Schurr J, B{\"u}rkel V and Kibble B~P 2009 {\em Metrologia\/} {\bf 46} 619

\bibitem{Ricketts:1988}
Ricketts B~W and Kemeny P~C 1988 {\em J. Phys. D: Appl. Phys.\/} {\bf 21} 483

\bibitem{Jeffery:1995}
Jeffery A, Elmquist R and Cage M 1995 {\em J. Res. Natl. Inst. Stan.\/} {\bf
  100} 677--685

\bibitem{Hartland:1995}
Hartland A, Kibble B, Rodgers P and Bohacek J 1995 {\em IEEE Trans. Instr.
  Meas.\/} {\bf 44} 245 --248

\bibitem{Chua:1999b}
Chua S~W, Hartland A and Kibble B 1999 {\em IEEE Trans. Instr. Meas.\/} {\bf
  48} 309 --313

\bibitem{Cage:1998a}
Cage M, Jeffery A, Elmquist R and Lee K 1998 {\em J. Res. Natl. Inst. Stan.\/}
  {\bf 103} 561--592

\bibitem{Sosso:1999}
Sosso A and Capra P~P 1999 {\em Rev. Sci. Instrum.\/} {\bf 70} 2082--2086

\bibitem{Sosso:2001}
Sosso A 2001 {\em IEEE Trans. Instr. Meas.\/} {\bf 50} 223--226

\bibitem{Schurr:2006}
Schurr J, Ahlers F~J, Hein G, Melcher J, Pierz K, Overney F and Wood B~M 2006
  {\em Metrologia\/} {\bf 43} 163
  \urlprefix\url{http://stacks.iop.org/0026-1394/43/i=1/a=021}

\bibitem{Schurr:2014}
Schurr J, Ahlers F and Pierz K 2014 {\em Metrologia\/} {\bf 51} 235--242

\bibitem{Ortolano:2012}
Ortolano M and Callegaro L 2012 {\em Metrologia\/} {\bf 49} 1

\bibitem{SpiceHome}
The {SPICE} page Online
  \urlprefix\url{http://bwrcs.eecs.berkeley.edu/Classes/IcBook/SPICE/}

\bibitem{Garg:1965}
Garg J~M and Carlin H~J 1965 {\em IEEE Trans. Circuit Theory\/}  59--73

\bibitem{Arnold:1982}
Arnold E 1982 {\em Surf. Sci.\/} {\bf 113} 239--243 ISSN 0039-6028
  \urlprefix\url{http://www.sciencedirect.com/science/article/pii/0039602882905921}

\bibitem{Popovic:1985}
Popovi\'c R~S 1985 {\em Solid-St. Electron.\/} {\bf 28} 711--716 ISSN 0038-1101
  \urlprefix\url{http://www.sciencedirect.com/science/article/pii/0038110185900218}

\bibitem{Salim:1992}
Salim A, Manku T, Nathan A and Kung W 1992 Modeling of magnetic-field sensitive
  devices using circuit simulation tools {\em Technical Digest of the 5th
  {IEEE} Solid-State Sensor and Actuator Workshop\/} pp 94--97

\bibitem{Salim:1995}
Salim A, Manku T and Nathan A 1995 {\em IEEE Trans. Comput.-Aided Design
  Integr. Circuits Syst.\/} {\bf 14} 464--469 ISSN 0278-0070

\bibitem{Chua:1980}
Chua L~O 1980 {\em IEEE Tran. Circuits Syst.\/} {\bf 27} 1014--1044

\bibitem{Willems:2010}
Willems J~C 2010 {\em IEEE Circuits Syst. Mag.\/} {\bf 10} 8--16

\bibitem{Belevitch:1968}
Belevitch V 1968 {\em Classical network theory\/} (San Francisco, CA, USA:
  Holden-Day)

\bibitem{Ihn:2010}
Ihn T 2010 {\em Semiconductor Nanostructures: Quantum States and Electronic
  Transport\/} (Oxford University Press)

\bibitem{CODATA:2010}
Mohr P~J, Taylor B~N and Newell D~B 2012 {\em Rev. Mod. Phys.\/} {\bf 84}
  1527--1605

\bibitem{Cage:1999a}
Cage M~E, Jeffery A and Matthews J 1999 {\em J. Res. Natl. Inst. Stan.\/} {\bf
  104} 529--556

\bibitem{Steer:2007}
Steer M~B 2007 {SPICE}: User's guide and reference Online
  \urlprefix\url{http://www.freeda.org/doc/SPICE/spice.pdf}

\bibitem{McAndrew:2005}
McAndrew C, Coram G, Blaum A and Pilloud O 2005 Correlated noise modeling and
  simulation {\em Technical Proceedings of the 2005 Workshop on Compact
  Modeling\/} (Anaheim, CA, US) pp 40--45

\bibitem{Rashid:2006}
Rashid M~H and Rashid H~M 2006 {\em SPICE for Power Electronics and Electric
  Power\/} 2nd ed (Boca Raton, FL, USA: CRC Press)

\bibitem{LTspice}
{Linear Technology} {LTspice IV}
  \urlprefix\url{http://www.linear.com/designtools/software/}

\bibitem{Ngspice}
{Ngspice} home page \urlprefix\url{http://ngspice.sourceforge.net/}

\bibitem{Delahaye:1993}
Delahaye F 1993 {\em J. Appl. Phys.\/} {\bf 73} 7914--7920

\bibitem{Ortolano:2015}
Ortolano M, Abrate M and Callegaro L 2015 {\em Metrologia\/} {\bf 52} 31

\bibitem{Tellegen:1948}
Tellegen B~D~H 1948 {\em Philips Res. Rept.\/} {\bf 3} 81--101

\bibitem{Viola:2014}
Viola G and {DiVincenzo} D~P 2014 {\em Phys. Rev. X\/} {\bf 4} 021019

\bibitem{Callegaro:2013}
Callegaro L, Ortolano M and Schurr J 2013 {\em Europhys. Lett.\/} {\bf 101}
  50003

\bibitem{LTspiceUG}
{LTspice IV} users' group \urlprefix\url{http://groups.yahoo.com/group/LTspice}

\bibitem{Shekel:1954}
Shekel J 1954 {\em Wireless Engineer\/} {\bf 31} 6--10

\bibitem{Roman:2010}
Roman S 2010 {\em Advanced Linear Algebra\/} 3rd ed (Springer)

\bibitem{Desoer:1969}
Desoer C~A and Kuh E~S 1969 {\em Basic circuit theory\/} (Singapore:
  McGraw-Hill, Inc.)

\bibitem{Chen:1991}
Chen W 1991 {\em Active network analysis\/} (Singapore: World Scientific
  Publishing Co.)

\bibitem{Twiss:1955}
Twiss R~Q 1955 {\em J. Appl. Phys.\/} {\bf 26} 599--602

\bibitem{Buttiker:1990}
B\"{u}ttiker M 1990 {\em Phys. Rev. Lett.\/} {\bf 65}(23) 2901--2904
  \urlprefix\url{http://link.aps.org/doi/10.1103/PhysRevLett.65.2901}

\bibitem{Buttiker:1992}
B\"{u}ttiker M 1992 {\em Phys. Rev. B\/} {\bf 46}(19) 12485--12507
  \urlprefix\url{http://link.aps.org/doi/10.1103/PhysRevB.46.12485}

\bibitem{Nyquist:1928}
Nyquist H 1928 {\em Phys. Rev.\/} {\bf 32}(1) 110--113

\bibitem{Landauer:1989}
Landauer R 1989 {\em Physica D\/} {\bf 38} 226--229

\bibitem{Priestley:1981}
Priestley M~B 1981 {\em Spectral analysis and time series\/} vol~2 (Academic
  Press)

\end{thebibliography}
\providecommand{\newblock}{}

\end{document}